\documentclass[prb,amsmath,amssymb,superscriptaddress,twocolumn]{revtex4}
\usepackage{amsmath}
\usepackage{braket}
\usepackage{graphicx}
\usepackage{multirow}
\usepackage{subfig}
\usepackage{bbm,color,ulem}
\usepackage{dsfont}
\allowdisplaybreaks

\begin{document}
\title{Simulation of Coupling Strength of Capacitively Coupled Singlet-Triplet Qubits}
\author{Donovan Buterakos}
\author{Robert E.\ Throckmorton}
\author{S.\ Das Sarma}
\affiliation{Condensed Matter Theory Center and Joint Quantum Institute, Department of Physics, University of Maryland, College Park, Maryland 20742-4111 USA}
\date{\today}

\begin{abstract}
We consider a system of two purely capacitively-coupled singlet-triplet qubits, and numerically simulate the energy structure of four electrons in two double quantum dots with a large potential barrier between them. We calculate the interqubit coupling strength using an extended Hund-Mulliken approach which includes excited orbitals in addition to the lowest energy orbital for each quantum dot. We show the coupling strength as a function of the qubit separation, as well as plotting it against the detunings of the two double quantum dots, and show that the general qualitative features of our results can be captured by a potential-independent toy model of the system.
\end{abstract}

\maketitle

\section{Introduction}

Semiconductor-based spin qubits are an attractive platform for quantum computing due to their long coherence times, fast gates, and potential for scalability.  Such qubits consist of one or more electrons trapped in quantum dots established near the surface of a semiconductor.  While such qubits have lower fidelities than competing platforms such as ion trap and superconducting qubits, much experimental progress has been made in recent years in improving the fidelity in semiconductor-based spin qubits (see, e.g., Ref.\ \onlinecite{NicholNPJQI2017}).  Several semiconductor-based qubit architectures have been proposed and studied both theoretically and experimentally, including the single spin qubit\cite{LossPRA1998,NowackScience2011,PlaNature2012,PlaNature2013,VeldhorstNatNano2014,BraakmanNatNano2013,OtsukaSciRep2016,ItoSciRep2016}, the singlet-triplet two-spin qubit\cite{LevyPRL2002,PettaScience2005,FolettiNatPhys2009,VanWeperenPRL2011,MauneNature2012,ShulmanScience2012,DialPRL2013,ShulmanNatCommun2014,ReedPRL2016,MartinsPRL2016}, the exchange-only three-spin qubit\cite{DiVincenzoNature2000,MedfordNatNano2013,MedfordPRL2013,EngSciAdv2015,ShimPRB2016}, and the hybrid qubit, consisting of three electrons in two quantum dots\cite{ShiPRL2012,KimNature2014,KimNPJQI2015}.  We will be focusing on the singlet-triplet qubit in this work.

A singlet-triplet qubit consists of two electrons confined to a double quantum dot, but free to tunnel between the two dots. Tunneling between the two dots creates an energy splitting $J$ between the singlet and triplet state, which is controlled by adjusting the detuning $\epsilon$, defined to be the potential energy difference between the two dots. In order to build a working quantum computer, we must first have a means to couple qubits so that multiqubit gates can be performed.  For singlet-triplet qubits, there are two such methods---capactive coupling and exchange coupling.  Capacitive coupling uses the fact that the singlet and triplet states have different electric dipole moments to realize a state-dependent coupling between two qubits, while exchange coupling simply uses an exchange interaction between one electron in one qubit and a neighboring electron in another qubit to couple the two.  While exchange coupling has the advantage of allowing the interqubit coupling to be tuned independently of the intraqubit exchange interactions, it has the disadvantage of enabling leakage of the qubits out of the computational singlet-triplet space.  Capacitive coupling, on the other hand, does not present the leakage problem, but the interqubit coupling is dependent on the intraqubit exchange couplings.  Our work will be dedicated to calculating the capacitive coupling within a simplified model of a pair of singlet-triplet qubits.

A number of previous works have considered microscopic models of one- and two-qubit systems of various types.  For example, Refs.\ \onlinecite{HiltunenPRB2014} and \onlinecite{SrinivasaPRB2015} employed tight-binding models, with the former concerning itself with determining the capacitive coupling within such a model.  Others have used harmonic potentials\cite{HuPRA2000,StepanenkoPRB2007,NielsenPRB2010,NielsenPRB2012,WhitePRB2018}.  While Ref.\ \onlinecite{NielsenPRB2012} concerned itself with determining the coupling between two singlet-triplet qubits, the relationship between the capacitive coupling and the intraqubit exchange couplings has not been explored, as we will do in this work.  It has often been assumed in both theoretical\cite{ButerakosPRB2018} and experimental\cite{NicholNPJQI2017,ShulmanScience2012} works that the capacitive coupling $J_{12}\propto J_1J_2$, where $J_1$ and $J_2$ are the intraqubit exchange couplings.  As this is just an empirical relation used in the experimental works, and can only be justified theoretically within a very crude approximation\cite{ButerakosPRB2018}, the assumptions of which are strongly violated in real experimental systems (i.e., the qubits are assumed to be very far apart in the argument, while the interqubit distance is comparable to the distance between dots within a qubit in real systems), this relation deserves further investigation.

The major features of a singlet-triplet qubit can be described by a Hund-Mulliken molecular orbit model\cite{BurkardPRB1999}. For zero detuning, the ground state is symmetric between the two dots, with one electron in each, which we will call the $(1,1)$ configuration, and the triplet state, which is spatially antisymmetric, carries only slightly more energy. As the detuning increases, the singlet state lowers its energy by mixing with the $(0,2)$ configuration, the state with both electrons in the lower energy dot. However, the triplet state is confined to the $(1,1)$ configuration due to the Pauli exclusion principle, and thus the exchange splitting can be increased by increasing the detuning. At a certain point, the energy difference between the two dots exceeds the energy cost due to the Coulomb repulsion, and it becomes energetically favorable for both electrons to occupy a single dot, if allowed by the spin state. At this point the singlet and triplet wave functions are relatively static with respect to changes in $\epsilon$, and $J$ simply grows linearly with $\epsilon$.

\begin{figure}
	\centering
		\includegraphics[width=\columnwidth]{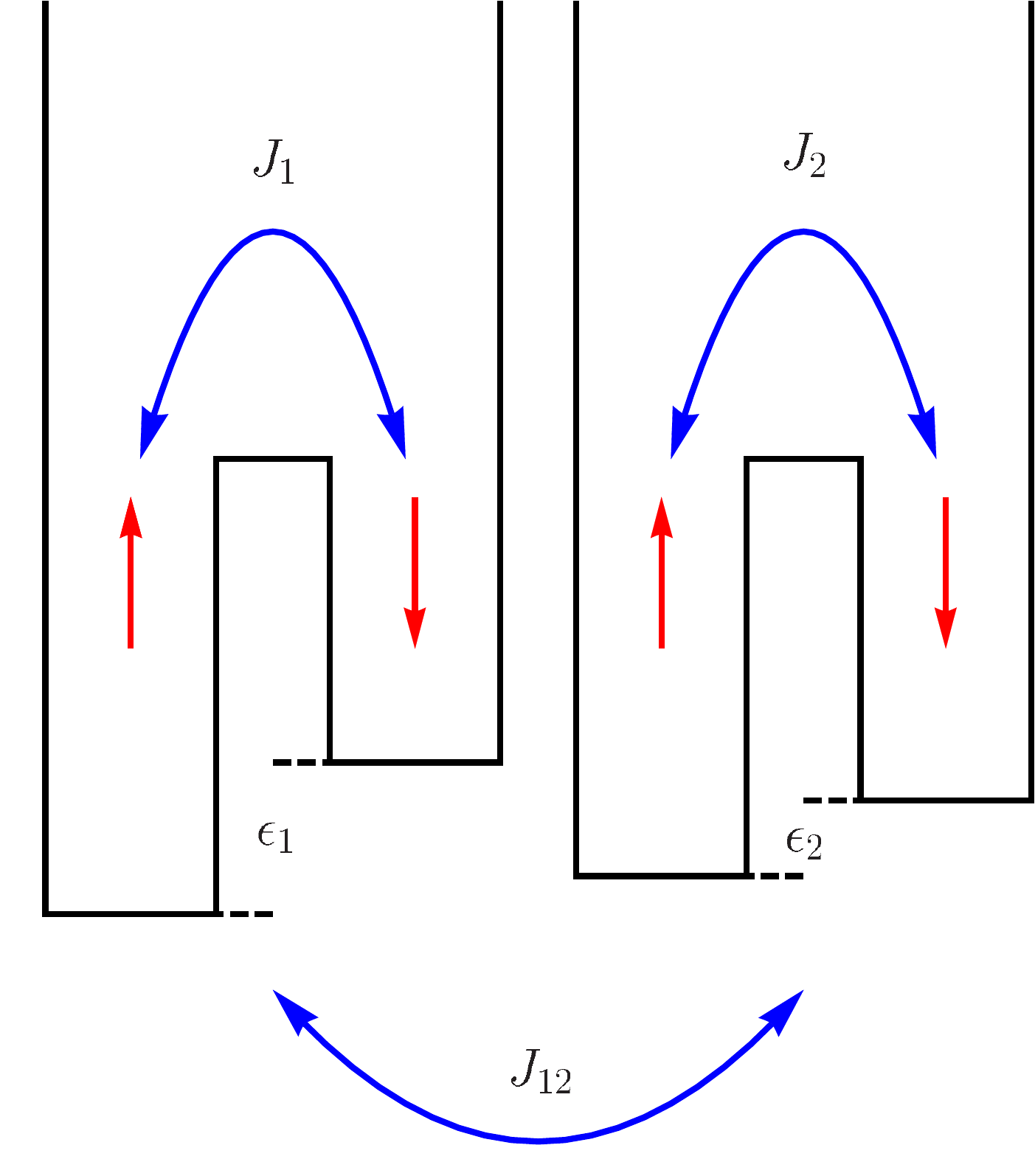}
	\caption{Schematic picture of the coupled double quantum dot qubit setup considered here, showing the intraqubit exchange couplings $J_1$ and $J_2$ and the interqubit coupling $J_{12}$.}
	\label{Fig:DQDIllust}
\end{figure}

We consider a system of two singlet-triplet qubits with an infinite potential barrier between them, so that the qubits interact only via the electron-electron Coulomb interaction. Because the singlet state is partially in the $(0,2)$ configuration, it has a dipole moment, whereas the dipole moment of the triplet state, which must stay in the $(1,1)$ configuration, is essentially zero. Thus for a system of two qubits, there exists a state-dependent dipole-dipole energy shift $J_{12}$ proportional (at least in the classical limit) to the strength of the two dipole moments.  We illustrate the basic setup that we will be considering, showing what all of the couplings represent, in Fig.\ \ref{Fig:DQDIllust}.

We numerically simulate a system of two capacitively coupled qubits in square-well quantum dot potentials. We extend the Hund-Mulliken model by considering the ground state and first excited states in each direction for each quantum dot, and construct the Hamiltonian by numerically evaluating two-body Coulomb integrals between each pair of states. We then numerically diagonalize the Hamiltonian, extracting the intraqubit exchange interaction strengths $J_1$ and $J_2$ and the interqubit coupling strength $J_{12}$ as functions of the detunings $\epsilon_1$ and $\epsilon_2$. We also find how $J_{12}$ varies as a function of qubit separation distance for fixed values of detuning. We gauge the precision of our results by calculating $J$ for a single qubit using up to four states in each dimension per dot, and comparing how the results change with an increasing number of basis states.

We then show that the general qualitative features of the results we obtain can be modeled using a simplified potential-independent Hund-Mulliken model, which depends only on the tunneling strength between the dots $t$, the detunings $\epsilon_1$ and $\epsilon_2$, and the classical dipole-dipole interaction energy $D$ which can be positive or negative depending on the qubit geometry and directions of bias for each qubit. We then analytically calculate $J_{12}$, and show that this closely matches the numerical results we obtain. Interestingly, we find that $J_{12}$ can only be approximated by the product of the dipole moments of two noninteracting qubits for very small detunings, when both qubits are in the (1,1) configuration. As the qubits are detuned away from this regime, the behavior of $J_{12}$ becomes more complicated. This model captures the qualitative features of the system, and is very general, in that the only dependence on a specific potential or geometry is in the parameters $t$ and $D$.

The rest of this paper is organized as follows. In Sec.\ II, we describe our numerical methods and provide our model Hamiltonian. In Sec.\ III present our numerical results, and in Sec.\ IV we derive the form of $J_{12}$ from the general, simplified Hund-Mulliken model for a two-qubit system. Finally, we give our conclusions in Sec.\ V.

\section{Numerical Methods}

A common method of modeling interactions between several quantum dots or other atom-like structures is a Hund-Mulliken approach, which has been used to model the dynamics of a single singlet-triplet qubit\cite{WhitePRB2018}. This approach involves constructing eigenstates of the interacting Hamiltonian from linear combinations of the lowest $n$ noninteracting orbitals of each quantum dot.

\begin{figure}
	\includegraphics[width=\columnwidth]{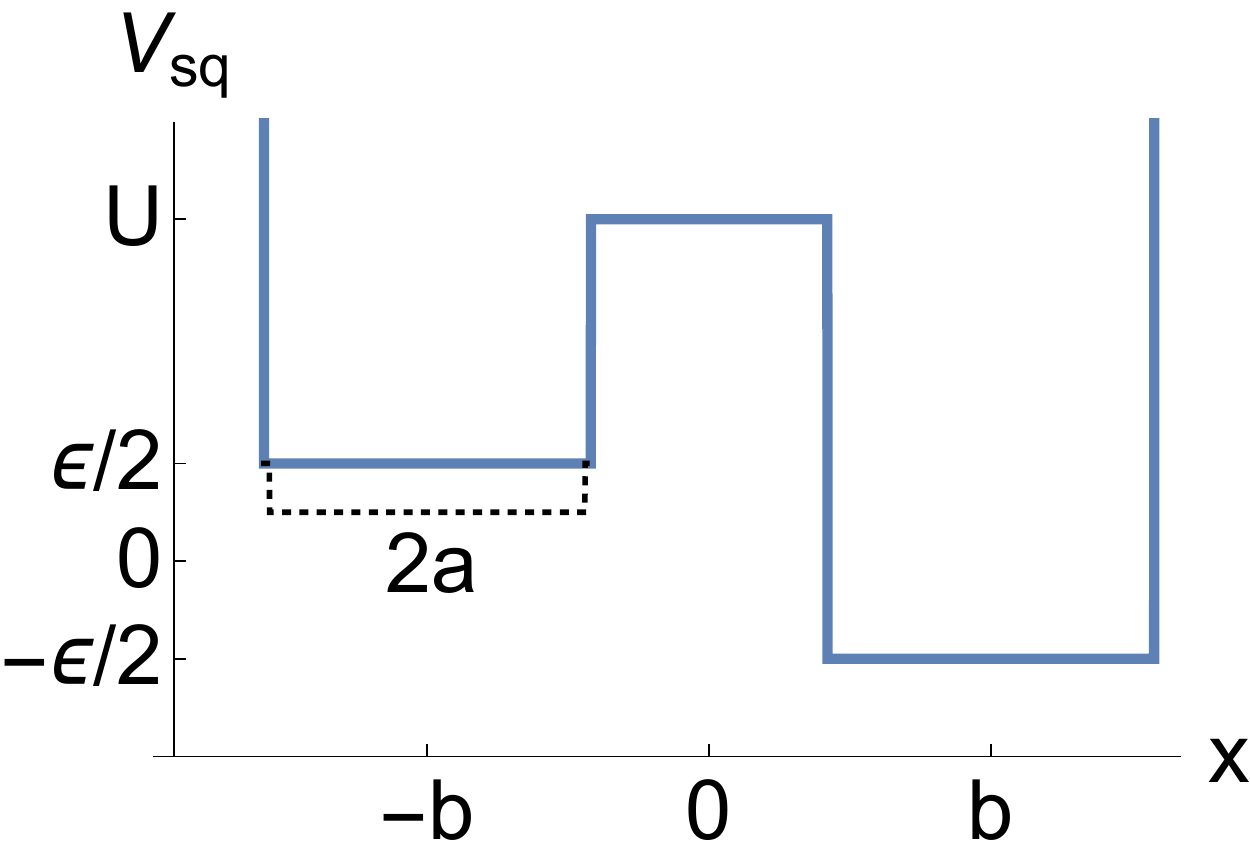}
	\caption{The effective square well potential for a double quantum dot defined by Eq.\ \eqref{eqn:veff} in the $x$ direction with detuning $\epsilon$, barrier height $U$, dot width $2a$, and dot spacing $2b$. The $z$ direction (not shown) forms a standard infinite square well with width $2a$.}
	\label{fig:vplot}
\end{figure}

We begin by presenting our model Hamiltonian.  The exact form of the Hamiltonian describing the systems studied in quantum dot experiments has not been derived from any microscopic model, and in fact may vary between different experimental implementations.  Therefore, our focus will still be on qualitative features and self-consistency rather than exact quantitative results and absolute accuracy. We first present our model for a single qubit, as the two-qubit model is easily generalized from it. We use a two-dimensional model, which we define to lie in the $xz$ plane, as the width of the dots in the $y$ direction is much smaller than the radius in the $xz$ plane. We take the $x$ axis to run parallel to the line connecting the two quantum dots. For simplicity, we restrict any applied magnetic field $\vec{B}$ to lie in the $xz$ plane; this allows us to choose the gauge in which the magnetic vector potential $\vec{A}=(xB_z-zB_x)\hat{y}$, so that the orbital effects of the magnetic field in the Hamiltonian are confined to the $y$ direction, allowing these orbital effects to be ignored for the purposes of our calculations.  We will also be working entirely within the subspace in which the electrons within a given qubit are in a singlet $\ket{S}=\frac{1}{\sqrt{2}}(\ket{\uparrow\downarrow}-\ket{\downarrow\uparrow})$ or triplet $\ket{T}=\frac{1}{\sqrt{2}}(\ket{\uparrow\downarrow}+\ket{\downarrow\uparrow})$ state, and thus there will be no magnetic field effects from spin either.  We therefore adopt the model,
\begin{equation}
H=\sum_i\left [\frac{p_i^2}{2m}+V(\vec{x}_i)\right ]+\frac{e^2}{\kappa|\vec{x}_1-\vec{x}_2|},
\label{eqn:singleqh}
\end{equation}
where $V(\vec{x}_i)$ is the single-particle potential defining the double quantum dot system.  We take $\kappa\approx 292$ for the purpose of making the Coulomb interaction a small perturbation; again, we are not interested in absolute accuracy, only in basic qualitative features, and in making our calculations as tractable as possible.  For reasons that we will explain shortly, we choose the following form for $V(\vec{x}_i)$:
\begin{equation}
V(\vec{x}_i)=V_{\text{sq}}(\vec{x_i})-V_{\text{corr}}(\vec{x_i}),
\label{eqn:v}
\end{equation}
where the square well potential $V_{\text{sq}}(\vec{x_i})$ is
\begin{equation}
V_{\text{sq}}(\vec{x})=
\begin{cases}
\epsilon/2 &\text{if } |z|<a\text{ and } {-b}-a<x<-b+a, \\
U &\text{if } |z|<a\text{ and } {-b}+a<x<b-a, \\
-\epsilon/2 &\text{if } |z|<a\text{ and\quad} b-a<x<b+a, \\
\infty &\text{otherwise},
\end{cases}
\label{eqn:veff}
\end{equation}
and the ``correction'' potential is
\begin{equation}
V_{\text{corr}}(\vec{x})=\begin{cases}
\int\frac{e^2}{\kappa|\vec{x}-\vec{x}'|}\psi_R^2(\vec{x}')\,d^2\vec{x}' &\text{if }x<0, \\
\int\frac{e^2}{\kappa|\vec{x}-\vec{x}'|}\psi_L^2(\vec{x}')\,d^2\vec{x}' &\text{if }x>0.
\end{cases}
\label{eqn:corr}
\end{equation}
The wave functions $\psi_{L/R}(\vec{x})$ are constructed by forming linear combinations of the lowest two eigenstates of $V_{\text{sq}}$ such that $\psi_{L/R}(\vec{x})=0$ at the midpoint of the potential barrier. The combination with most of its ``weight'' in the left dot is $\psi_L(\vec{x})$, and that with most of its ``weight'' in the right dot is $\psi_R(\vec{x})$.  These are essentially the ground states of the potentials obtained by placing an infinite wall at the midpoint of the double quantum dot system.  This ``correction'' potential is similar to a mean field potential, such as in Ref.\ \onlinecite{HuPRA2000}, and in fact becomes identical to it in the limit of an infinitely high barrier.  Since the integrals defining this potential can be difficult to evaluate, we numerically calculate the first several terms in a multipole expansion for the given charge distribution.  Here, $\epsilon$ is the detuning, $U$ is the barrier height, $a$ is the dot radius, and $b$ is half the separation between dots.  We provide a plot of $V_{\text{sq}}(\vec{x})$ along the $x$ axis for $|z|<a$ in Fig.\ \ref{fig:vplot}.

As the Schr\"odinger equation for a multielectron wave function cannot be exactly solved, a common method of modeling such systems involves choosing a finite basis $\ket{\Psi_i}$, calculating the projection of the Hamiltonian onto this basis,
\begin{equation}
H_\Psi=\sum_{ij}\ket{\Psi_i}\braket{\Psi_i|H|\Psi_j}\bra{\Psi_j},
\end{equation}
and numerically diagonalizing $H_\Psi$.  As the size of the basis $\ket{\Psi_i}$ grows, $H_\Psi$ approaches the exact Hamiltonian $H$, and thus its eigenvalues converge to the exact energies. The standard Hund-Mulliken model corresponds to choosing $\ket{\Psi_i}$ to be the set of antisymmetrized two-electron products of single electron states $\ket{\psi_i}$, where $\ket{\psi_i}$ includes the ground states of the two quantum dots tensored with the up and down spin states, under the restriction that $S_z\ket{\Psi_i}=0$ (i.e., restricted to the singlet-triplet computational space). We extend this model by allowing $\ket{\psi_i}$ to include the ground state orbital and first few excited orbitals of each quantum dot. In order to help this method to converge quickly, we attempt to maximize the overlap of the basis states with the true wave function by accounting for the mean-field effects of other electrons in our choice of basis. Rather than choosing the basis to be eigenstates of the bare potential $V$, we use eigenstates of the effective potential formed by adding to the bare potential the mean field correction $V_{\text{corr}}$ defined in Eq.\ \eqref{eqn:corr}. We define this ``effective'' potential to be $V_{\text{sq}}$, and thus the bare potential is formed by subtracting the mean-field correction from $V_{\text{sq}}$, as in Eq.\ \eqref{eqn:v}. In effect, we are assuming that the gates defining the quantum dots are tuned in such a way as to produce a potential that, when the mean field experienced by one electron due to the other is added to it, will produce a semi-infinite square well on each side of the double quantum dot system.

Two choices for the potential are commonly used in theoretical work---a polynomial that forms an (approximate) harmonic potential at each of the two dots, and the square well potential that we use in this work. We choose the square well potential for two reasons.  First, screening due to the metal gates causes the potential inside the dot to become nearly flat, and thus the square well potential is physically closer to the true potential. One downside of this choice is that, in order to keep the potential separable in the $x$ and $z$ directions (for ease of computation), we must use the square dots produced by the potential in Eq.\ \eqref{eqn:veff} rather than circular dots; however, this should only have a relatively small effect on the exchange coupling.  Secondly, the exchange coupling depends very heavily on the tunneling coefficient between the two dots, and thus it is essential for the single particle states to accurately match the potential both inside and between the two quantum dots.  Eigenstates for electrons in potentials given by high order polynomials cannot be found exactly, and, while using Gaussian wave functions will accurately represent the wave function inside the quantum dots, they do not accurately approximate the magnitude between the two dots, and thus we require a large number of such basis states to obtain accurate results for the exchange energy. In contrast, the eigenstates for a square well potential are trivial to calculate, and thus the tunneling behavior of the wave function can be encoded in the basis states themselves. Simulations for the harmonic potential have also been done, for one\cite{HuPRA2000,NielsenPRB2010} and two\cite{NielsenPRB2012,HiltunenPRB2014} qubits, but we find a square well to be more convenient for this context.

To model a system of two capacitively-coupled qubits, we simply take two copies of the single qubit Hamiltonian given by Eq.\ \eqref{eqn:singleqh} and add interqubit Coulomb interaction terms, as follows:
\begin{equation}
H_I+H_{II}+\sum_{i\in I}\sum_{j\in II}\frac{e^2}{\kappa|\vec{x}_i-\vec{x}_j|}.
\end{equation}
We assume that the two qubits are a distance $2c$ apart, measured between the midpoints of their closest dots.  In this case, $V_{\text{corr}}$ is modified to take into account the electrons in the other qubit as well. It should be understood that the square well parts of the potentials are finite within each qubit, and infinite elsewhere, so that there is no tunneling between the qubits.  As a result, the qubits are coupled purely capacitively---there is no exchange coupling between them.

\section{Numerical results}

\begin{figure}
	\includegraphics[width=\columnwidth]{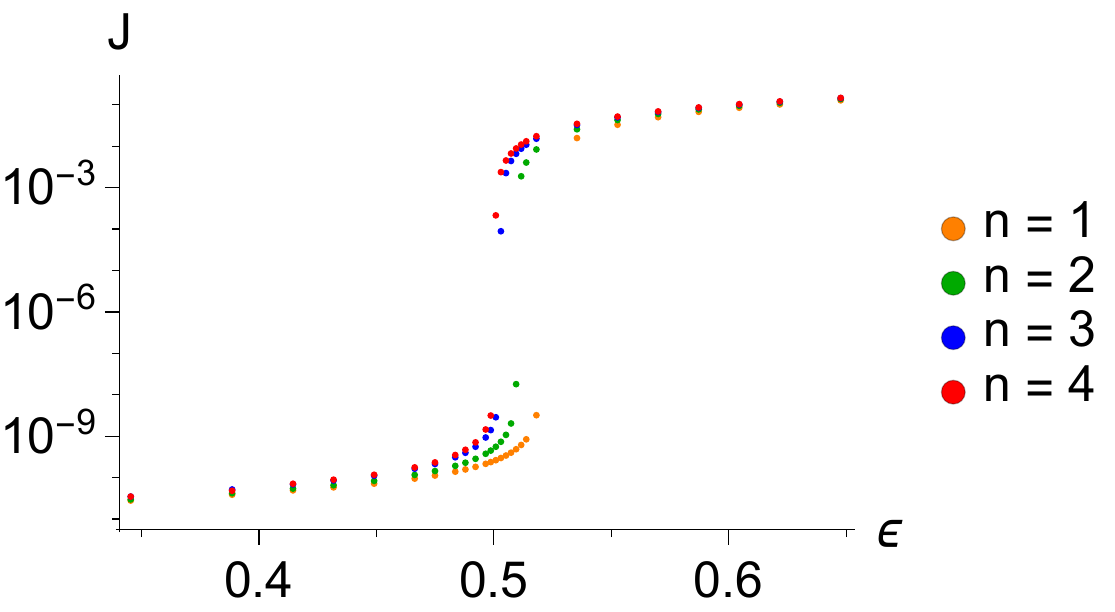}
	\caption{Plot of the exchange coupling $J$ for $U=17.4\mathcal{E}_0$ as a function of the detuning $\epsilon$ for a single qubit. For increasing values of $n$, the shape of the plot remains similar, with only the position of the transition point shifting left or right slightly.}
	\label{fig:onequbit}
\end{figure}

Matrix elements of the Hamiltonian in the basis of the tensor products of single-electron eigenstates of $V_{\text{sq}}$ are calculated via numerical integration, with the two-body Coulomb integrals being done in Fourier space in order to reduce the dimensionality of the integrals.  We choose a cutoff $n$, corresponding to the maximum number of orbitals we consider for each dot in each direction ($n=1$ corresponds to using only the ground state square well orbitals of each dot, $n=2$ allows orbitals with one excitation in either the $x$ or $z$ direction, and so on). We give energies in terms of the ground state energy of a single square well in one dimension of radius $a$, which we will denote $\mathcal{E}_0=\frac{\pi^2}{2m(2a)^2}$. For our simulation, we use the relative dot size $a/b=0.58$ and $c=b$ resulting in all four dots being evenly spaced, matching the architecture in Ref.\ \onlinecite{ZajacPRAPP2016}. We adjust the barrier height $U$ so that it is larger than the highest energy state in the basis $\ket{\psi_i}$, in order that no oscillatory behavior is present between the dots, as this would require many additional states to cancel out.

\begin{figure*}
	\centering
	\subfloat[][Plot of the capacitive coupling $J_{12}$ as a function of the detunings $\epsilon_1$ and $\epsilon_2$ for $D<0$.]{
		\includegraphics[width=.9\columnwidth]{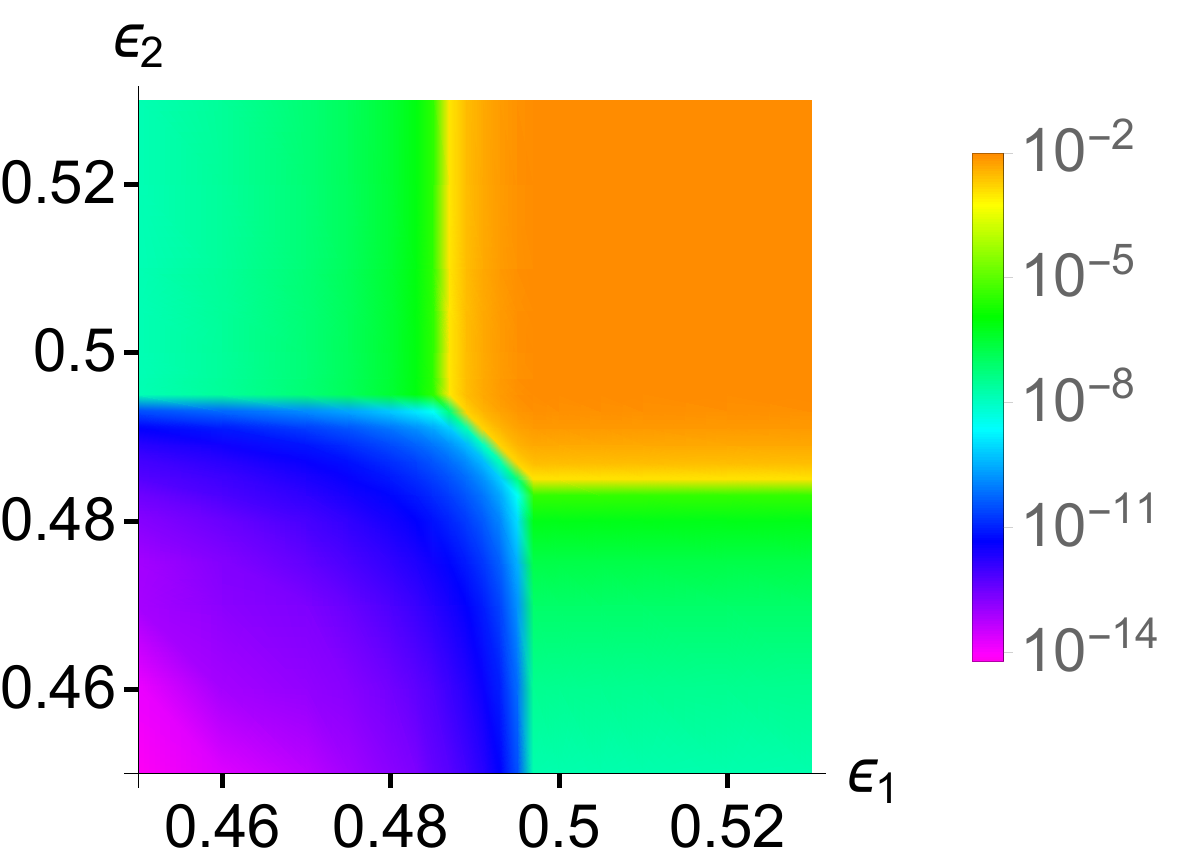}
		\label{fig:j12negd}
	}
	\qquad
	\subfloat[][Plot of the capacitive coupling $J_{12}$ as a function of the detunings $\epsilon_1$ and $\epsilon_2$ for $D>0$.]{
		\includegraphics[width=.9\columnwidth]{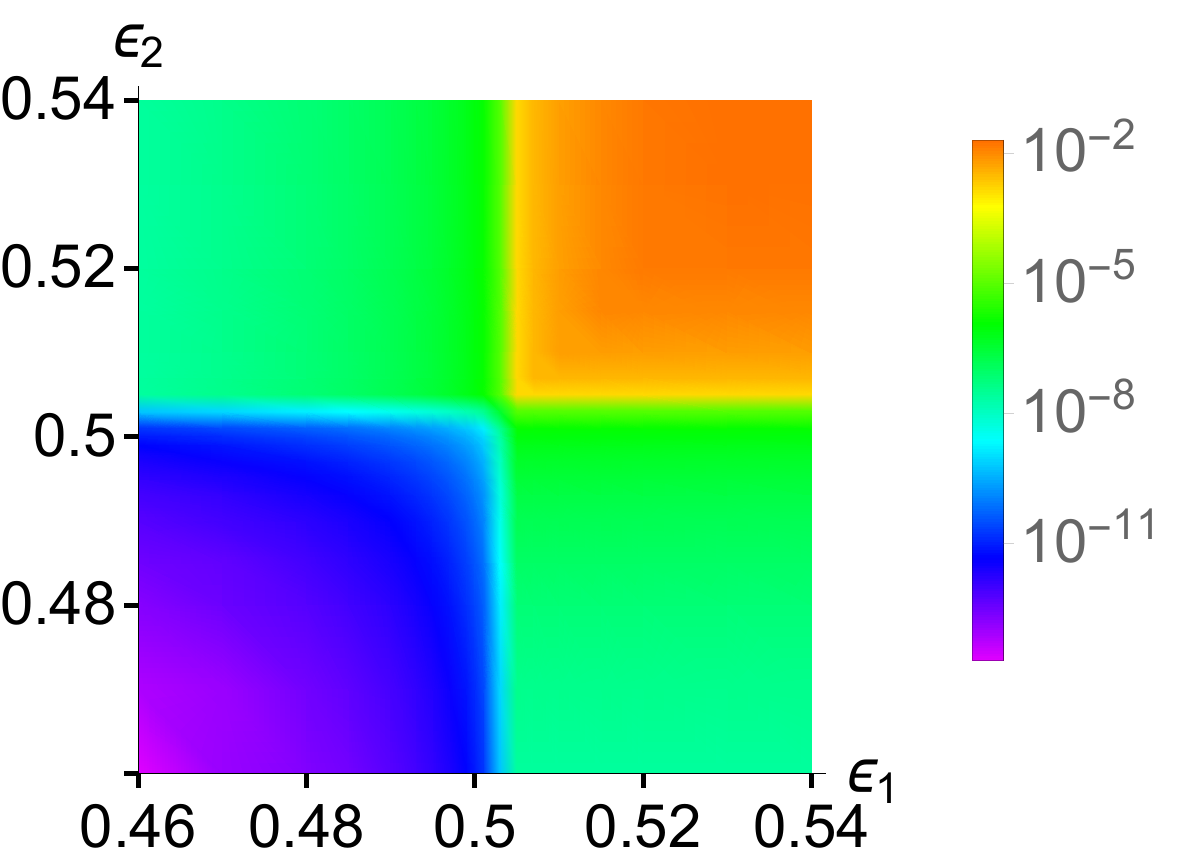}
		\label{fig:j12posd}
	}\\
	\subfloat[][Plot of the ratio $J_{12}/(J_1J_2)$ for $D<0$.]{
		\includegraphics[width=.9\columnwidth]{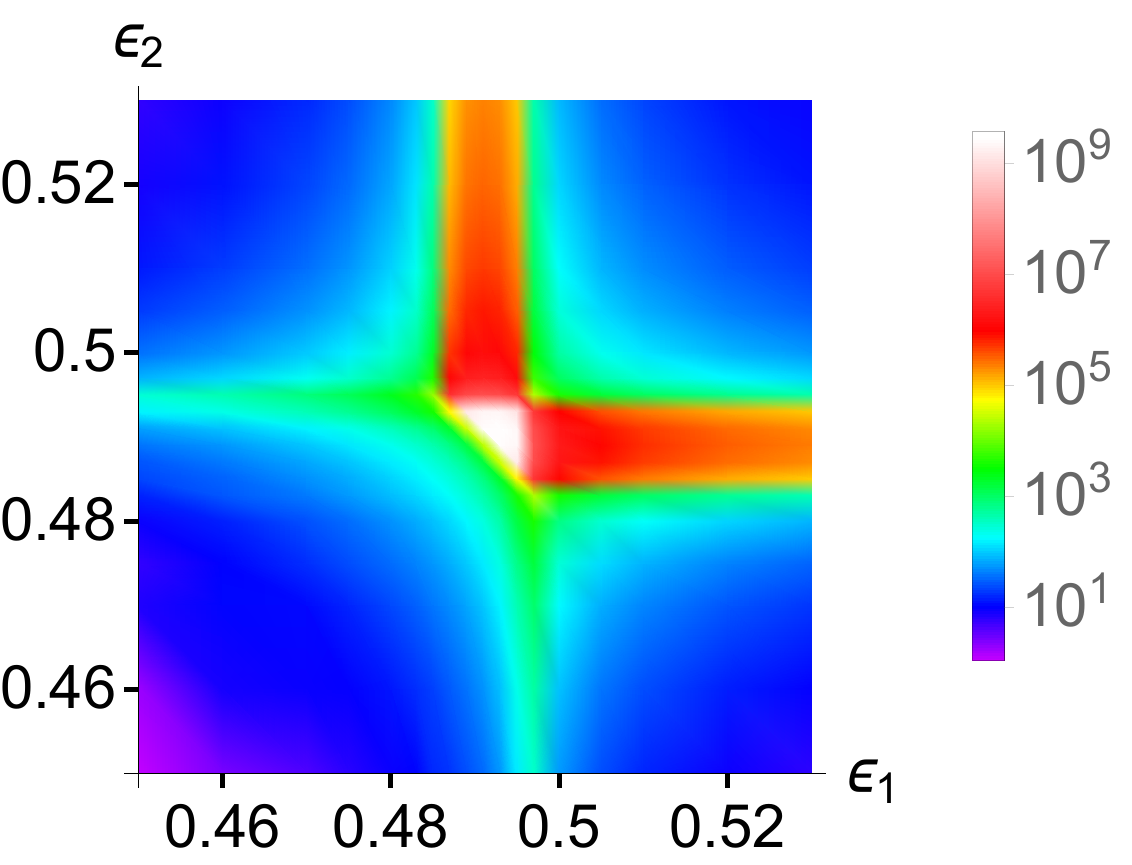}
		\label{fig:ratnegd}
	}
	\qquad
	\subfloat[][Plot of the ratio $J_{12}/(J_1J_2)$ for $D>0$.]{
		\includegraphics[width=.9\columnwidth]{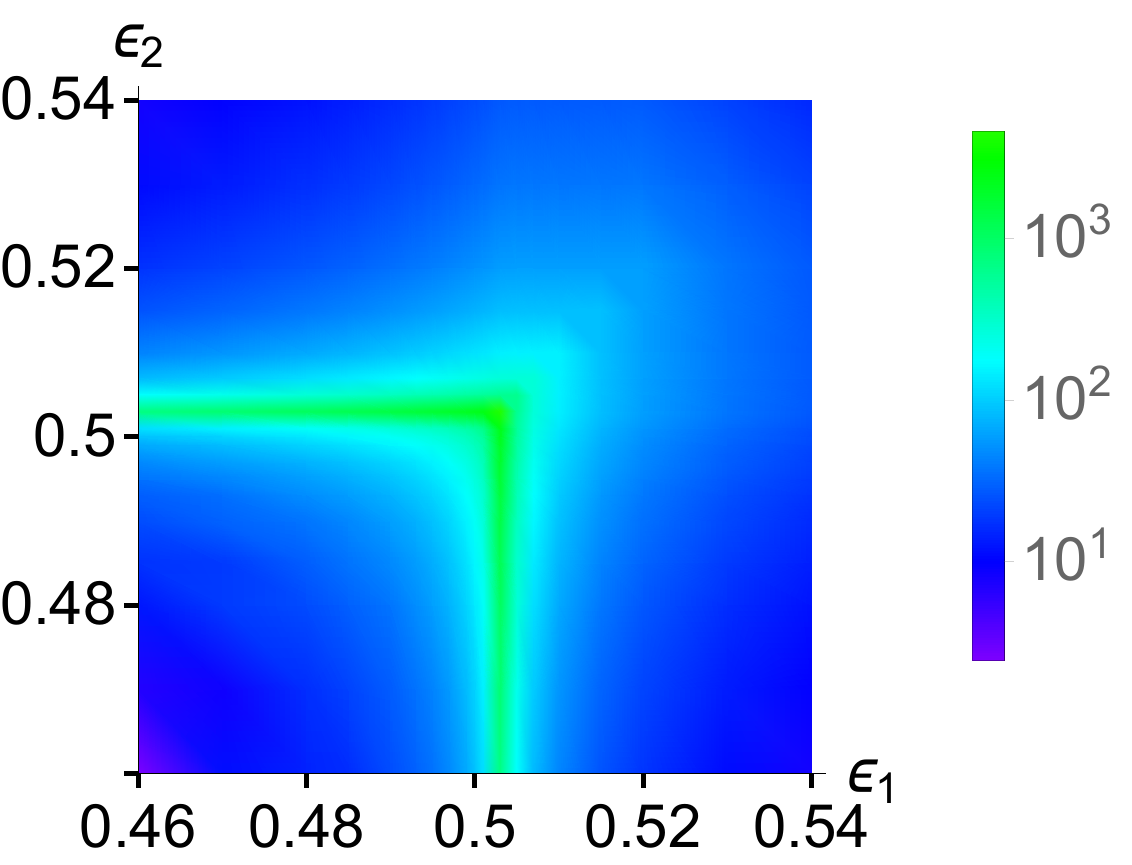}
		\label{fig:ratposd}
	}
	\caption{Plots of $J_{12}$ (top row), and the ratio $J_{12}/(J_1J_2)$ (bottom row) for  negative dipole-dipole interaction energy $D<0$ (left column) and positive dipole-dipole interaction energy $D>0$ (right column) as functions of $\epsilon_1$ and $\epsilon_2$. For these plots, $U=10\mathcal{E}_0$.  Taken together, these show that, for the model that we are considering, the relationship $J_{12}\propto J_1J_2$ holds only approximately.}
	\label{fig:results}
\end{figure*}

We diagonalize $H_{\Psi}$ for a single qubit for cutoffs of $n=1$ through $4$ and extract the value of the exchange coupling $J$ by subtracting the energy of the lowest-lying triplet state from that of the lowest-lying singlet state.  We plot this exchange coupling as a function of the detuning $\epsilon$ for several values of $n$ in Fig.\ \ref{fig:onequbit}.  We see that this method converges very quickly; the qualitative features remain the same even for a small basis size.  We note that $J$ increases relatively slowly as a function of $\epsilon$ except near the critical value around $\epsilon\approx 0.5$ which is the value for which we observe a transition between the $(1,1)$ state and the $(0,2)$ state.  We see that this transition point shifts only slightly when extending the basis size, illustrating that our numerical methods converge quickly as a function of the basis size.


We now consider the case of two capacitively coupled singlet-triplet qubits.  We again construct the full four-electron basis states from the single-electron eigenstates, under the assumption that there are two electrons in each qubit, restricting ourselves to those states for which the total $z$ component of the electrons' spins $S_z=0$, and in which each qubit is in the singlet-triplet computational subspace (i.e., the spin state of the electrons is not, say, $\ket{\uparrow\uparrow\downarrow\downarrow}$).  We then extract the values of the exchange couplings $J_1$ and $J_2$ and the capacitive coupling $J_{12}$ by identifying the lowest-energy states with the spin configurations, $\ket{SS}$, $\ket{ST}$, $\ket{TS}$, and $\ket{TT}$, and fitting the resulting diagonalized effective Hamiltonian to the form: 
\begin{equation}
H=\tfrac{1}{2}J_1Z_1+\tfrac{1}{2}J_2Z_2+\tfrac{1}{4}J_{12}(Z_1-1)(Z_2-1)+K
\label{eqn:hlogical}
\end{equation}

where $Z_i$ is the logical Pauli $Z$ matrix acting on qubit $i$, and $K$ is an (unimportant) constant. We do this twice, once for the case where the detunings $\epsilon_1$ and $\epsilon_2$ are in the same direction, (making the dipole moments parallel, and thus giving a negative interaction energy), and once for the case in which the detunings are in opposite directions (thus giving a positive interaction energy). We provide plots of our numerical results for $U=17.4\mathcal{E}_0$ as a function of the detunings $\epsilon_1$ and $\epsilon_2$ in Fig.\ \ref{fig:results}. Finally, we fix the detunings and adjust the distance between qubits $2c$, in order to show how $J_{12}$ depends on qubit separation. We find that for large distances, $J_{12}$ falls off as would a classical dipole-dipole interaction. However, as $c$ becomes smaller, $J_{12}$ plateaus rather than rising indefinitely.

\begin{figure}
	\includegraphics[width=\columnwidth]{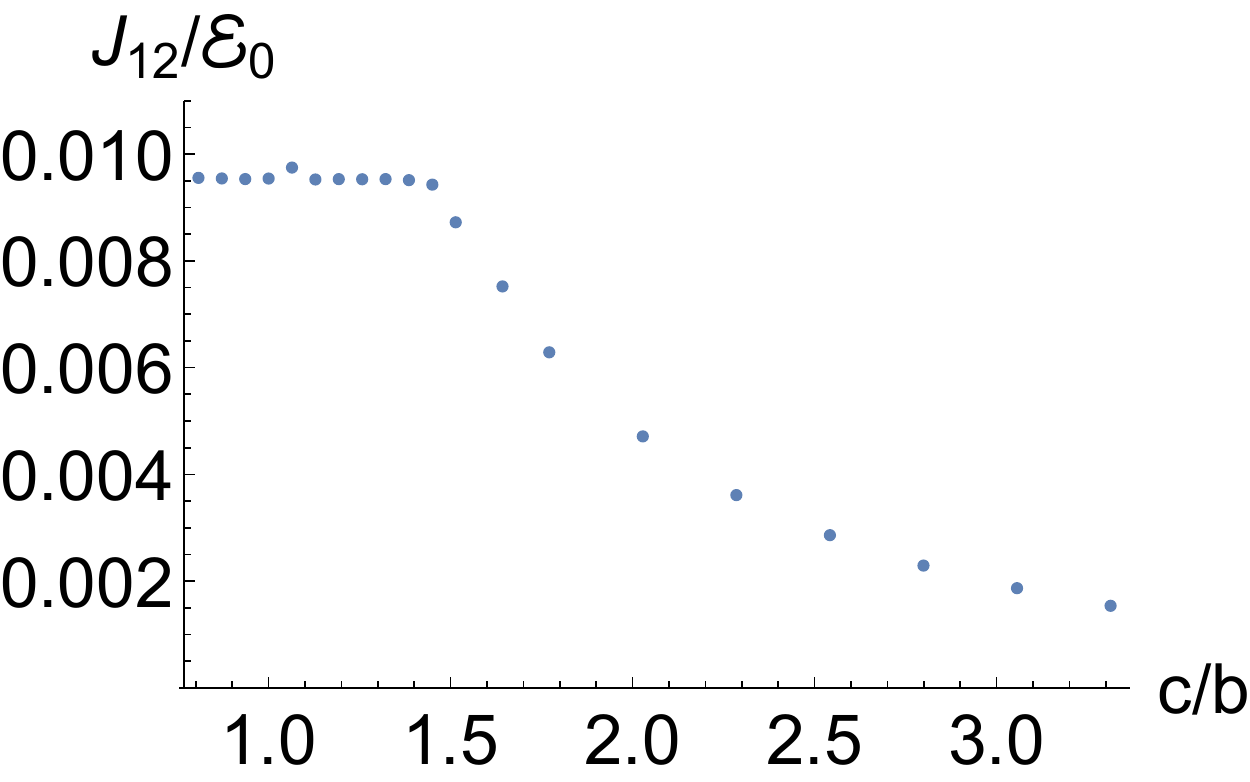}
	\caption{Plot of $J_{12}$ versus half-distance between qubits $c$ for $\epsilon_1=\epsilon_2=.51\mathcal{E}_0$.}
	\label{fig:dist}
\end{figure}

\section{Discussion}

We find that a four-dot potential-independent Hund-Mulliken model can accurately describe the behavior of the results we obtain. For each qubit, we use the standard Hund-Mulliken model of two detuned quantum dots, considering only the lowest energy orbital of each dot. The potential can be characterized by only two parameters: the detuning $\epsilon$, defined to be the energy difference between the lowest energy orbitals of the two dots; and the tunneling coefficient $t$, given by the term in the Hamiltonian which mixes the two states. Considering two electrons bound to these two states, by the Pauli exclusion principle, the triplet state can only consist of one electron in each dot. The singlet state can occupy one of three electron configurations: $(0,2)$, $(1,1)$, and $(2,0)$. Since the detuning will always favor one dot, we neglect the singlet state with both electrons in the higher energy dot, (for example the $(2,0)$ state). The Hamiltonian for the singlet state is then given by:
\begin{equation}
H_S=
\begin{pmatrix}
0&t\\t&-\epsilon'
\end{pmatrix}
\end{equation}
in the $(S_{11},S_{02})$ basis, where $\epsilon'=\epsilon-C$ is the energy difference between the $S_{11}$ and $S_{02}$ states, which depends on the difference in Coulomb repulsion energy $C$ and the bare detuning $\epsilon$. With this definition, the energy of the triplet state is 0, and thus the exchange splitting $J$ is simply the absolute value of the singlet ground state energy:
\begin{equation}
E_S=-J=-\frac{\epsilon'}{2}-\sqrt{\Big(\frac{\epsilon'}{2}\Big)^2+t^2}
\label{eqn:jdef}
\end{equation}

To find $J_{12}$, we consider two copies of this single-qubit system, and add a dipole-dipole interaction term. Since we are concerned with the capacitive coupling strength, we shall assume that the two qubits are separated by an infinite potential barrier, so that there is no tunneling between the two qubits and hence no interqubit exchange coupling. This means that the qubits act as if they were isolated, except that the total energy is changed when both qubits possess a dipole moment. Since the dipole moment of the triplet state is zero, the $\ket{ST}$, $\ket{TS}$, and $\ket{TT}$ states are all unaffected, and are treated as above. The $\ket{SS}$ state can be modeled by taking two copies of the single-qubit Hamiltonian $H_S$ and adding a constant dipole-dipole interaction energy $D$ to the $\ket{S_{02}S_{02}}$ state, as done in Ref.\ \onlinecite{SrinivasaPRB2015}, yielding the following Hamiltonian:
\begin{align}
H_{SS}&=H_S(\epsilon_1)\otimes\mathds{1}+\mathds{1}\otimes H_S(\epsilon_2)+\frac{D}{4}(\sigma_z-\mathds{1})\otimes(\sigma_z-\mathds{1})\nonumber\\
&=\begin{pmatrix}
0&t&t&0\\t&-\epsilon_2'&0&t\\t&0&-\epsilon_1'&t\\0&t&t&-\epsilon_1'-\epsilon_2'+D
\end{pmatrix}
\label{eqn:model}
\end{align}

For a system of two coupled singlet-triplet qubits, $J_1$, $J_2$, and $J_{12}$ are defined such that the Hamiltonian in the logical subspace, that is, the space spanned by the lowest energy states of each spin configuration $(TT,TS,ST,SS)$, as given by Eq.\ \eqref{eqn:hlogical}. Using this definition, the values of $J_1$ and $J_2$ remain the same as given in Eq.\ \eqref{eqn:jdef}, and the capacitive coupling strength is given by:
\begin{equation}
J_{12}=E_{SS}+J_1+J_2
\label{eqn:j12def}
\end{equation}
where $E_{SS}$ is the lowest energy eigenvalue of $H_{SS}$ defined above. It is possible to represent $E_{SS}$ completely algebraically, as the zero of the fourth degree characteristic polynomial of $H_{SS}$, but the resulting expression is unwieldy, so we extract the general behavior of the system by looking at several limits. First, we look at the behavior of $E_{SS}$ in the limit where $t\rightarrow0$. In this case $E_{SS}$ is given by:
\begin{equation}
E_{SS}=\min\big(0,\ -\epsilon_2',\ -\epsilon_1',\ -\epsilon_1'-\epsilon_2'+D\big)
\label{eqn:esst0}
\end{equation}

\begin{figure}[h]
	\includegraphics[width=.7\columnwidth]{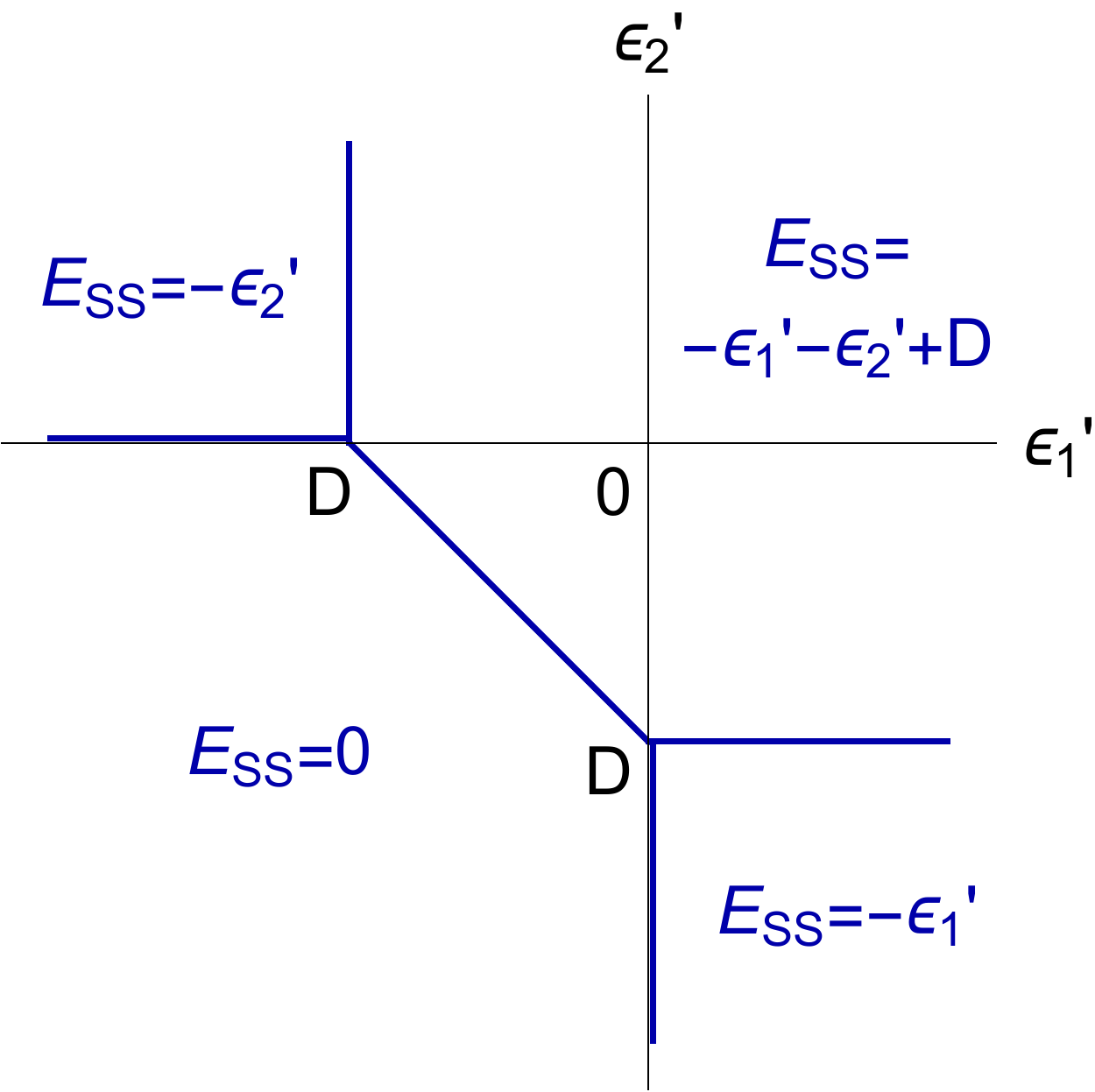}
	\includegraphics[width=.7\columnwidth]{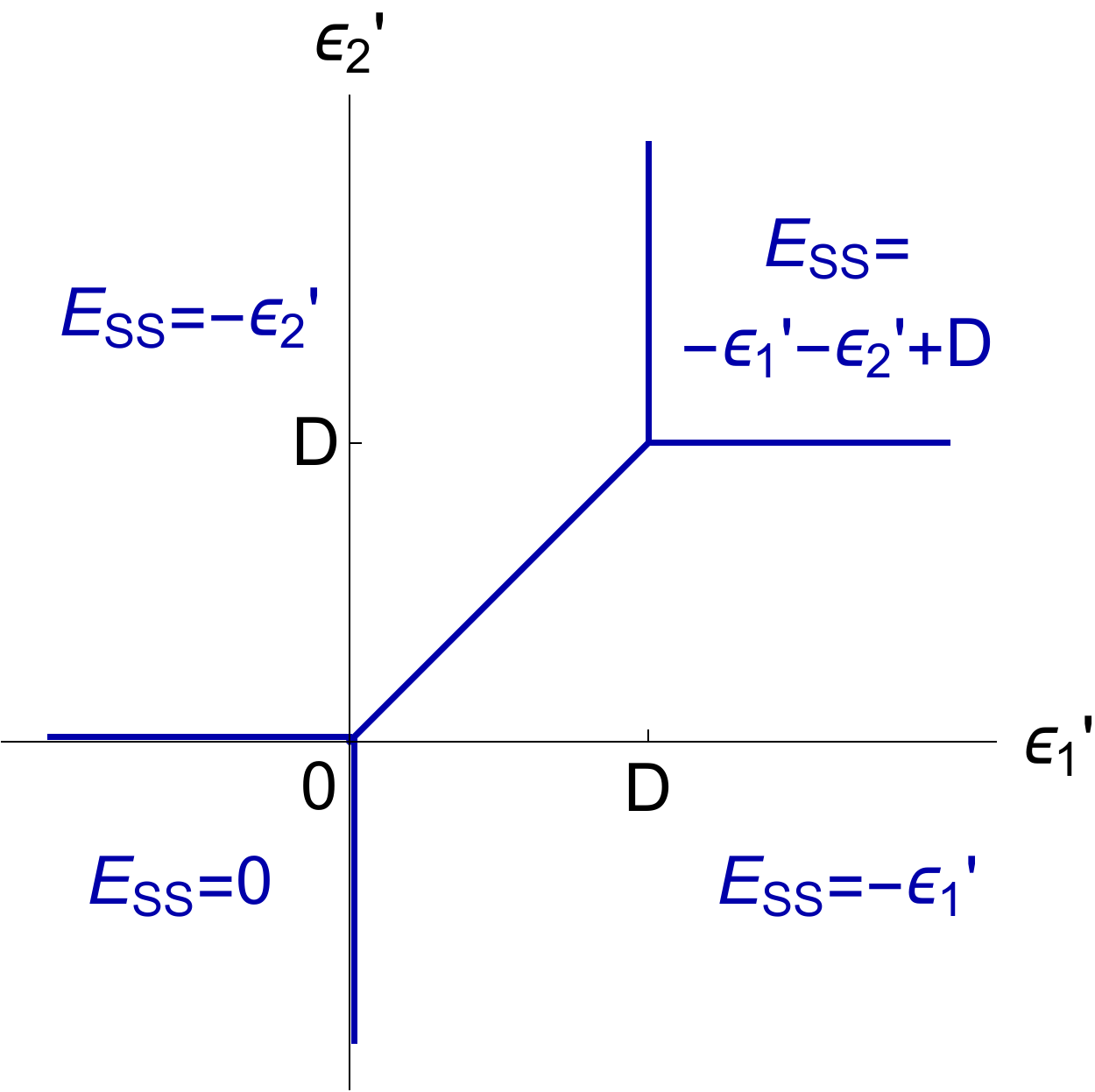}
	\caption{A diagram showing $E_{SS}$ as a function of $\epsilon_1'$ and $\epsilon_2'$ for $t=0$, which is given by Eq.\ \eqref{eqn:esst0}, for $D<0$ (top) and $D>0$ (bottom).}
	\label{fig:essplot}
\end{figure} 

This divides the $\epsilon_1\epsilon_2$-plane into four regions, distinguished by which of the four eigenvalues is smallest in each region, as shown in Fig.\ \ref{fig:essplot}. On scales much larger than $t$, the transitions between regions appear to be sharp corners, which then become smooth avoided crossings on scales comparable to $t$. As shown, the shape of the regions varies depending on whether $D$ is greater than or less than zero, and the behavior of the system is significantly different in these two cases. When $D$ is negative, there is an extra incentive for both qubits to be in the $(0,2)$ configuration, and so the region characterized by the $S_{02}S_{02}$ configuration (the top-right region of Fig.\ \ref{fig:essplot}) becomes larger, which is seen in our results in Fig.\ \ref{fig:j12negd}. Notably, this means that if one qubit (for example, qubit 2) is in or near the $(0,2)$ configuration, the other qubit transitions at a lower detuning than normal, since the dipole-dipole attraction energy helps to offset the intraqubit Coulomb repulsion energy. Thus in this region (near the top of Fig.\ \ref{fig:essplot}), $J_{12}$ and $J_1$ transition at different points, with $|J_{12}|$ becoming large while $J_1$ is still nearly zero. Conversely, when $D$ is positive, an additional energy cost must be paid to force both qubits into the $(0,2)$ configuration simultaneously, meaning the transition for $E_{SS}$ requires a larger detuning. However, $J_1$ and $J_2$ are unaffected by $D$, and still transition at $\epsilon_1'$ or $\epsilon_2'=0$ respectively, and thus for $0>\epsilon_i'>D$, $J_{12}$ is dominated by the $J_1+J_2$ terms.

We can perform an expansion of the exact eigenvalues of Eq.\ \eqref{eqn:model}. In the limit where $J_2\ll t$, which corresponds to $-\epsilon_2'\gg t,|D|$, we find the following form for $J_{12}$:
\begin{equation}
J_{12}=\frac{DJ_1^2J_2^2}{t^2(J_1^2+t^2)}+O\Big[J_2^3\Big]
\end{equation}

Thus we see that for small $J_1$ or $J_2$, the coupling strength acts like the product of the dipole moments of two isolated qubits. When $J_2\gg t$, which corresponds to $\epsilon_2'\gg t,|D|$, we find:
\begin{equation}
J_{12}=\frac{D}{2}+\sqrt{\Big(\frac{\epsilon_1'}{2}\Big)^2+t^2}-\sqrt{\Big(\frac{\epsilon_1'-D}{2}\Big)^2+t^2}
\end{equation}

For $D<0$, $J_{12}$ transitions from near zero behavior to linear growth in $\epsilon_1'$ at the point $\epsilon_1'=D$, then transitions from linear growth to a constant, $D$, at $\epsilon_1'=0$. The case where $D>0$ is similar, except that the initial transition from near zero to linear behavior occurs at $\epsilon_1'=0$, and the transition from linear to constant happens at $\epsilon_1'=D$. Interestingly, since the transition between the $(1,1)$ and $(0,2)$ configurations happens at the second transition point when $D>0$, the coupling strength $J_{12}$ reaches close to its maximum value while qubit 1 is still almost entirely in the $(1,1)$ configuration. In comparison, for $D<0$, $|J_{12}|$ only begins to grow quickly after both qubits have transitioned to the $(0,2)$ configuration.

This model explains the behavior of the coupling strength $J_{12}$ as a function of distance between qubits. The dipole-dipole interaction energy $D$ decreases with increasing distance. We plot a region where $\epsilon_1',\epsilon_2',D>0$, and so for large $D$, $J_{12}$ simply falls off with increasing distance as well. However, when $D$ is made large enough, the values of detunings chosen transition from the region characterized by $E_{SS}\approx-\epsilon_1'-\epsilon_2'+D$ to the regions $-\epsilon_1'$ and $-\epsilon_2'$. Since the energy of these regions does not depend on $D$ to lowest order, $J_{12}$ should be constant for large $D$.

\section{Conclusion}

We simulated a system of two capacitively coupled singlet-triplet qubits, and numerically calculated the coupling strength $J_{12}$ using an extended Hund-Mulliken approach, where we numerically diagonalize the projection of the Hamiltonian onto a finite subspace. We then show that our results can be approximated by a generic minimal model.

In our simulation, we assume for simplicity that the qubits may be modeled as 2D square wells, with a potential barrier of width $b$ and height $U$ separating two semi-finite square wells with width $2a$, all bounded by infinite walls, along the $x$ direction and a simple infinite square well of width $2a$ in the $z$ direction.  We also add detuning $\epsilon$, defined as the difference of the potential in the two dots.  While this choice is motivated somewhat by experiment, it is not an exact model---we are only interested in general qualitative features, not in a precise simulation of actual experimental systems.  Carefully choosing the subspace allows for a quickly converging energy spectrum to be calculated, from which we determine the values of $J_1$, $J_2$, and $J_{12}$.  We do not expect that a more detailed analysis, using a more realistic model, would produce results much different from those presented in this work.  If we considered a fully three-dimensional system, for example, giving our system a small thickness along the $y$ direction, then this only introduces additional very high-energy states in comparison to those that we are concerned with in our system.  Even if our numerical results may differ from what is seen in actual experimental systems, our results should still serve as a useful guide for experiments, showing what general features one should expect to find in actual systems.

Our generic model is independent of the qubit geometry and potential, in that these only affect the model through the values of two constant parameters: the tunneling strength $t$, and the dipole-dipole interaction energy $D$. Our model consists of two pairs of quantum dots, where tunneling is allowed between the two dots within of each pair, but where some infinite potential barrier separates the two pairs, so that there is no tunneling from one pair to another. This ensures that there is no exchange coupling between the qubits, as it is our goal to examine only the capacitive coupling strength. We define the detunings $\epsilon_1$ and $\epsilon_2$ to be the energy difference between dots in the first and second pair respectively, and adjust these detunings by the energy cost of having two electrons in a single dot, so that $\epsilon_i'=0$ at the $(1,1)$ to $(0,2)$ transition point of an isolated qubit. We construct a Hamiltonian over the space spanned by the lowest energy states of each dot, and from this Hamiltonian determine the behavior of $J_{12}$.

For a system of two qubits, a naive approach is to treat $J_{12}$ as the product of two isolated single-qubit dipole moments; however, we find that this only holds far from the transition points. Close to the transition, more complex structure appears, and is dependent on the sign of $D$. For $D<0$, we find a shift in the $J_{12}$ transition boundary, so that the singlet-singlet state transitions from the $(1,1)$ configuration to $(0,2)$ at a lower detuning than does the singlet-triplet state. This disparity means that the system cannot be quantitatively treated as a product of two individual dipole moments, and is evidenced in our numerics by the fact that the ratio $\frac{J_{12}}{J_1J_2}$ varies by over 6 orders of magnitude in the vicinity of the transition point. For $D>0$, we find a similar behavior, where the singlet-singlet transitions at a higher detuning than the singlet-triplet state, which causes $J_{12}$ to grow very large while one qubit still remains in the $(1,1)$ configuration. Thus, a more quantitatively accurate relationship than a simple dipole-dipole model is needed for many applications.  For example, our previous work on error correction in coupled singlet-triplet qubits\cite{ButerakosPRB2018,ButerakosPRB2018_2} assumes this proportionality, and presents tables of parameters defining pulses which dynamically correct crosstalk errors under this model. Since the basic proportionality relationship used for $J_{12}$ differs from realistic systems, the exact pulses we initially presented will not be applicable. However, the method presented for deriving these pulses is quite general, and can be used to generate new pulse sequences using a more accurate system-dependent model for the strength of $J_{12}$. In fact, our dynamical decoupling technique developed in Refs. \onlinecite{ButerakosPRB2018} and \onlinecite{ButerakosPRB2018_2} for correcting ST qubit errors would still apply formally exactly in the same manner, but the detailed pulses would have to be tailored to the specific ST inter-qubit coupling (which no longer can be assumed to be proportional to $J_1J_2$ in all of the parameter space) relevant for the specific experimental system by first carrying out calculations along the line of what is presented in the current work.  This would obviously make realistic quantum error correction rather costly in terms of computational demand, but we see no way out of such detailed numerical calculations if one is serious about developing fault-tolerant semiconductor spin qubits in the future.  Error correction depends crucially on a knowledge of the system Hamiltonian. We do, however, find that the proportionality of $J_{12}$ to $J_1J_2$ holds over a wide range of parameters, away from transitions of one or both qubits from a $(1,1)$ state to a $(0,2)$ state, or vice versa.

\acknowledgments
This work is supported by the Laboratory for Physical Sciences.


\begin{thebibliography}{99}
\bibitem{NicholNPJQI2017}J.\ M.\ Nichol, L.\ A.\ Orona, S.\ P.\ Harvey, S.\ Fallahi, G.\ C.\ Gardner, M.\ J.\ Manfra, and A.\ Yacoby, npj Quant.\ Inf.\ {\bf 3}, 3 (2017).
\bibitem{LossPRA1998}D.\ Loss and D.\ P.\ DiVincenzo, \pra {\bf 57}, 120 (1998).
\bibitem{NowackScience2011}K.\ Nowack, M.\ Shaffei, M.\ Laforest, G.\ E.\ D.\ K.\ Prawiroatmodjo, L.\ R.\ Schreiber, C.\ Reichl, W.\ Wegscheider, and L.\ M.\ K.\ Vandersypen, Science {\bf 333}, 1269 (2011).
\bibitem{PlaNature2012}J.\ J.\ Pla, K.\ Y.\ Tan, J.\ P.\ Dehollain, W.\ H.\ Lim, J.\ J.\ L.\ Morton, D.\ N.\ Jamieson, A.\ S.\ Dzurak, and A.\ Morello, Nature (London) {\bf 489}, 541 (2012).
\bibitem{PlaNature2013}J.\ J.\ Pla, K.\ Y.\ Tan, J.\ P.\ Dehollain, W.\ H.\ Lim, J.\ J.\ L.\ Morton, F.\ A.\ Zwanenburg, D.\ N.\ Jamieson, A.\ S.\ Dzurak, and A.\ Morello, Nature (London) {\bf 496}, 334 (2013).
\bibitem{VeldhorstNatNano2014}M.\ Veldhorst, J.\ C.\ C.\ Hwang, C.\ H.\ Yang, A.\ W.\ Leenstra, B.\ de Ronde, J.\ P.\ Dehollain, J.\ T.\ Muhonen, F.\ E.\ Hudson, K.\ M.\ Itoh, A.\ Morello, and A.\ S.\ Dzurak, Nat.
Nanotechnol. 9, 981 (2014).
\bibitem{BraakmanNatNano2013}F.\ R.\ Braakman, P.\ Barthelemy, C.\ Reichl, W.\ Wegscheider, and L.\ M.\ K.\ Vandersypen, Nat. Nanotechnol. 8, {\bf 432} (2013).
\bibitem{OtsukaSciRep2016}T.\ Otsuka, T.\ Nakajima, M.\ R.\ Delbecq, S.\ Amaha, J.\ Yoneda, K.\ Takeda, G.\ Allison, T.\ Ito, R.\ Sugawara, A.\ Noiri, A.\ Ludwig, A.\ D.\ Wieck, and S.\ Tarucha, Sci. Rep. {\bf 6}, 31820 (2016).
\bibitem{ItoSciRep2016}T.\ Ito, T.\ Otsuka, S.\ Amaha, M.\ R.\ Delbecq, T.\ Nakajima, J.\ Yoneda, K.\ Takeda, G.\ Allison, A.\ Noiri, K.\ Kawasaki, and S.\ Tarucha, Sci. Rep. {\bf 6}, 39113 (2016).
\bibitem{LevyPRL2002}J.\ Levy, \prl {\bf 89}, 147902 (2002).
\bibitem{PettaScience2005}J.\ Petta, A.\ Johnson, J.\ Taylor, E.\ Laird, A.\ Yacoby, M.\ Lukin, C.\ Marcus, M.\ Hanson, and A.\ Gossard, Science {\bf 309}, 2180 (2005).
\bibitem{FolettiNatPhys2009}S.\ Foletti, H.\ Bluhm, D.\ Mahalu, V.\ Umansky, and A.\ Yacoby, Nat. Phys. {\bf 5}, 903 (2009).
\bibitem{VanWeperenPRL2011}I.\ van Weperen, B.\ D.\ Armstrong, E.\ A.\ Laird, J.\ Medford, C.\ M.\ Marcus, M.\ P.\ Hanson, and A.\ C.\ Gossard, \prl {\bf 107}, 030506 (2011).
\bibitem{MauneNature2012}B.\ M.\ Maune, M.\ G.\ Borselli, B.\ Huang, T.\ D.\ Ladd, P.\ W.\ Deelman, K.\ S.\ Holabird, A.\ A.\ Kiselev, I.\ Alvarado-Rodriguez, R.\ S.\ Ross, A.\ E.\ Schmitz, M.\ Sokolich, C.\ A.\ Watson, M.\ F.\ Gyure, and A.\ T.\ Hunter, Nature (London) {\bf 481}, 344 (2012).
\bibitem{ShulmanScience2012}M.\ D.\ Shulman, O.\ E.\ Dial, S.\ P.\ Harvey, H.\ Bluhm, V.\ Umansky, and A.\ Yacoby, Science {\bf 336}, 202 (2012).
\bibitem{DialPRL2013}O.\ E.\ Dial, M.\ D.\ Shulman, S.\ P.\ Harvey, H.\ Bluhm, V.\ Umansky, and A.\ Yacoby, \prl {\bf 110}, 146804 (2013).
\bibitem{ShulmanNatCommun2014}M.\ D.\ Shulman, S.\ P.\ Harvey, J.\ M.\ Nichol, S.\ D.\ Bartlett, A.\ C.\ Doherty, V.\ Umansky, and A.\ Yacoby, Nat. Commun. {\bf 5}, 5156 (2014).
\bibitem{ReedPRL2016}M.\ D.\ Reed, B.\ M.\ Maune, R.\ W.\ Andrews, M.\ G.\ Borselli, K.\ Eng, M.\ P.\ Jura, A.\ A.\ Kiselev, T.\ D.\ Ladd, S.\ T.\ Merkel, I.\ Milosavljevic, E.\ J.\ Pritchett, M.\ T.\ Rakher, R.\ S.\ Ross, A.\ E.\ Schmitz, A.\ Smith, J.\ A.\ Wright, M.\ F.\ Gyure, and A.\ T.\ Hunter, \prl {\bf 116}, 110402 (2016).
\bibitem{MartinsPRL2016}F.\ Martins, F.\ K.\ Malinowski, P.\ D.\ Nissen, E.\ Barnes, S.\ Fallahi, G.\ C.\ Gardner, M.\ J.\ Manfra, C.\ M.\ Marcus, and F.\ Kuemmeth, \prl {\bf 116}, 116801 (2016).
\bibitem{DiVincenzoNature2000}D.\ P.\ DiVincenzo, D.\ Bacon, J.\ Kempe, G.\ Burkard, and K.\ B.\ Whaley, Nature (London) {\bf 408}, 339 (2000).
\bibitem{MedfordNatNano2013}J. Medford, J. Beil, J. M. Taylor, S. D. Bartlett, A. C. Doherty, E. I. Rashba, D. P. DiVincenzo, H. Lu, A. C. Gossard, and C. M. Marcus, Nat. Nanotechnol. {\bf 8}, 654 (2013).
\bibitem{MedfordPRL2013}J.\ Medford, J.\ Beil, J.\ M.\ Taylor, E.\ I.\ Rashba, H.\ Lu, A.\ C.\ Gossard, and C.\ M.\ Marcus, \prl {\bf 111}, 050501 (2013).
\bibitem{EngSciAdv2015}K.\ Eng, T.\ D.\ Ladd, A.\ Smith, M.\ G.\ Borselli, A.\ A.\ Kiselev, B.\ H.\ Fong, K.\ S.\ Holabird, T.\ M.\ Hazard, B.\ Huang, P.\ W.\ Deelman, I.\ Milosavljevic, A.\ E.\ Schmitz, R.\ S.\ Ross, M.\ F.\ Gyure, and A.\ T.\ Hunter, Sci. Adv. {\bf 1}, e1500214 (2015).
\bibitem{ShimPRB2016}Y.-P.\ Shim and C.\ Tahan, \prb {\bf 93}, 121410(R) (2016).
\bibitem{ShiPRL2012}Z.\ Shi, C.\ B.\ Simmons, J.\ R.\ Prance, J.\ K.\ Gamble, T.\ S.\ Koh, Y.-P.\ Shim, X.\ Hu, D.\ E.\ Savage, M.\ G.\ Lagally, M.\ A.\ Eriksson, M.\ Friesen, and S.\ N.\ Coppersmith, \prl {\bf 108}, 140503 (2012).
\bibitem{KimNature2014}D.\ Kim, Z.\ Shi, C.\ B.\ Simmons, D.\ R.\ Ward, J.\ R.\ Prance, T.\ S.\ Koh, J.\ K.\ Gamble, D.\ E.\ Savage, M.\ G.\ Lagally, M.\ Friesen, S.\ N.\ Coppersmith, and M.\ A.\ Eriksson, Nature (London) {\bf 511}, 70 (2014).
\bibitem{KimNPJQI2015}D.\ Kim, D.\ R.\ Ward, C.\ B.\ Simmons, D.\ E.\ Savage, M.\ G.\ Lagally, M.\ Friesen, S.\ N.\ Coppersmith, and M.\ A.\ Eriksson, npj Quant.\ Inf.\ {\bf 1}, 15004 (2015).
\bibitem{HiltunenPRB2014}T.\ Hiltunen and A.\ Harju, \prb {\bf 90}, 125303 (2014).
\bibitem{SrinivasaPRB2015}V.\ Srinivasa and J.\ M.\ Taylor, \prb {\bf 92}, 235301 (2015).
\bibitem{HuPRA2000}X.\ Hu and S.\ Das Sarma, \pra {\bf 61}, 062301 (2000).
\bibitem{StepanenkoPRB2007}D.\ Stepanenko and G.\ Burkard, \prb {\bf 75}, 085324 (2007).
\bibitem{NielsenPRB2010}E.\ Nielsen, R.\ W.\ Young, R.\ P.\ Muller, and M.\ S.\ Carroll, \prb {\bf 82}, 075319 (2010).
\bibitem{NielsenPRB2012}E.\ Nielsen, R.\ P.\ Muller, and M.\ S.\ Carroll, \prb {\bf 85}, 035319 (2012).
\bibitem{WhitePRB2018}Z.\ White and G.\ Ramon, \prb {\bf 97}, 045306 (2018).
\bibitem{ButerakosPRB2018}D.\ Buterakos, R.\ E.\ Throckmorton, and S.\ Das Sarma, \prb {\bf 97}, 045431 (2018).
\bibitem{BurkardPRB1999}G.\ Burkard, D.\ Loss, and D.\ P.\ DiVincenzo \prb {\bf 59}, 2070 (1999).
\bibitem{ZajacPRAPP2016} D.\ M.\ Zajac, T.\ M.\ Hazard, X.\ Mi, E.\ Nielsen, and J.\ R.\ Petta, Phys. Rev. App. {\bf 6}, 054013 (2016).
\bibitem{ButerakosPRB2018_2}D.\ Buterakos, R.\ E.\ Throckmorton, and S.\ Das Sarma, \prb {\bf 98}, 035406 (2018).
\end{thebibliography}
\end{document}